\documentclass[nofootinbib,prd,aps,superscriptaddress,preprintnumbers]{revtex4}
\pdfoutput=1
\usepackage{amsmath,amssymb,euscript}
\usepackage{color}
\usepackage{accents}
\usepackage{hyperref}
\usepackage{ulem}
\usepackage{epsfig}
\usepackage{varioref}
\usepackage{xcolor}
\usepackage{verbatim}

\renewcommand{\arraystretch}{1.2} 

\def\beq{\begin{equation}}
\def\eeq{\end{equation}}

\renewcommand{\emph}{\textit}

\graphicspath{{figs/}}

\begin{document}

\begin{flushright}
CERN-TH-2017-017
\end{flushright}

\title{Tau properties in $B\to D\tau\nu$ from visible final-state kinematics}

\author{Rodrigo Alonso}
\affiliation{CERN, Theoretical Physics Department, CH-1211 Geneva 23, Switzerland}
\author{Jorge Martin Camalich}
\affiliation{CERN, Theoretical Physics Department, CH-1211 Geneva 23, Switzerland}
\author{Susanne Westhoff}
\affiliation{Institute for Theoretical Physics, Heidelberg University, D-69120 Heidelberg, Germany}

\vspace{1.0cm}
\begin{abstract}
\vspace{0.2cm}\noindent 
In semi-leptonic $B$ decays with a tau lepton, features of the production process are imprinted on the tau helicity states. Since the tau momentum cannot be fully reconstructed experimentally, the available information on the tau properties is encoded in its visible decay products. Focusing on the process $B\to D\tau\nu$, we find explicit relations between the tau properties and the kinematics of the charged particles in the decays $\tau\to\pi\nu$, $\tau\to\rho\nu$, and $\tau\to\ell\nu\bar{\nu}$. In particular, we show that the perpendicular polarization, $P_\perp$, and the forward-backward asymmetry, $A_\tau$, of the tau lepton can simultaneously be extracted from an angular asymmetry of the charged particle against the $D$ meson. For the most sensitive decay channel, $\tau\to\pi\nu$, we expect a relative statistical precision of about $10\%$ for $P_\perp$ and $A_\tau$ in a measurement based on $50\,$ab$^{-1}$ of data at BELLE II.
\end{abstract}
\maketitle

%%%%%%%%%%%%%%%%%%%%%%%%%%%%%%%%%%%%%%%%%%%%%%%%%%%%%%%%%%%%%%%%%%%%%%%%%%%%%%

\section{Introduction}
The tau leptons in $B\to D\tau\nu$ and $B\to D^{\ast}\tau\nu$ decays serve as a wide test ground for physics in and beyond the standard model (SM). Within the SM, these decays differ from the decays $B\to D^{(\ast)}\ell\nu$ with light leptons $\ell=e,\mu$ by the larger mass of the tau lepton, which facilitates its production through a longitudinally polarized virtual $W$ boson~\cite{Korner:1989qb}. The ratios of decay rates into taus and light leptons, $R_D$ and $R_{D^{\ast}}$, can be predicted very precisely using model-independent calculations of form factors and experimental input from the spectra of $B\to D\ell\nu$ decays~\cite{Isgur:1989ed,Isgur:1989vq,Manohar:2000dt,deRafael:1993ib,Boyd:1995sq,Boyd:1997kz,Caprini:1997mu,Lattice:2015rga,Na:2015kha}.
On the experimental side, increasingly precise measurements of total decay rates and single differential distributions have been achieved by BaBar~\cite{Lees:2012xj,Lees:2013uzd}, BELLE~\cite{Huschle:2015rga,Sato:2016svk,Abdesselam:2016xqt}, and LHCb~\cite{Aaij:2015yra} (see Ref.~\cite{Amhis:2016xyh} for a review). The large amount of data expected in the near future from BELLE II and LHCb will enable us to probe $B\to D^{(\ast)}\tau\nu$ decays in great detail. Exploring the properties of $B$ decays with tau leptons is thus a key target at future $B$ physics experiments~\cite{Aushev:2010bq}.

The precise SM predictions of $B\to D^{(\ast)}\tau\nu$ observables are also key to searches for physics beyond the SM. Two decades ago, semi-leptonic $B$ decays with taus had already been proposed as sensitive probes of new charged scalars~\cite{Tanaka:1994ay,Kiers:1997zt} and for a model-independent analysis of new physics~\cite{Goldberger:1999yh}. More recently, an observed discrepancy between SM predictions and measurements of $R_D$ and $R_{D^{\ast}}$~\cite{Amhis:2016xyh} has triggered a major effort to scrutinize possible explanations in terms of new physics~(see e.g. Refs.~\cite{Tanaka:2012nw,Crivellin:2012ye,Freytsis:2015qca,Bardhan:2016uhr,Bhattacharya:2016zcw,Alonso:2016oyd,Celis:2016azn} for an overview of models and model-independent analyses). This unsolved puzzle calls for a careful investigation of other observables, such as those related to tau polarizations~\cite{Tanaka:1994ay}, angular distributions of the tau~\cite{Duraisamy:2013kcw,Becirevic:2016hea}, or other decay modes induced by the same elementary $b\to c\tau\nu$ transition~\cite{Detmold:2015aaa,Lytle:2016ixw}.

Nonetheless, an important experimental challenge of $B$ decays with tau leptons is their fast decay. The final state necessarily involves one or more neutrinos, which escape the detector and hinder a full reconstruction of the tau kinematics.~\footnote{A detection of the displaced vertex of the tau decay could add this missing piece of information~\cite{Tanaka:1994ay}.} The maximal accessible information on the $b\to c\tau\nu$ transition is encoded in the visible decay products of the $\tau$ lepton, where the three dominant decay modes $\tau\to\ell\nu\bar{\nu}$, $\tau\to\rho\nu$, and $\tau\to\pi\nu$ make up a branching ratio of more than $70\%$. It is therefore suitable to construct observables directly from final-state kinematics  of the visible decay particle $d=\{\ell,\rho,\pi\}$, without relying on the reconstruction of the tau momentum. This approach has been considered~\cite{Kiers:1997zt} and later pioneered~\cite{Nierste:2008qe} in the context of searches for charged scalars in $B\to D\tau\nu$. Subsequently, a variety of observables -- partially or fully based on different visible final states -- have been proposed to measure tau polarizations~\cite{Tanaka:2010se,Ivanov:2017mrj}, angular asymmetries~\cite{Sakaki:2012ft}, CP violation~\cite{Hagiwara:2014tsa}, to study the impact of leptonic tau decay on light lepton distributions~\cite{Bordone:2016tex}, or to facilitate a comprehensive analysis of new physics~\cite{Alonso:2016gym,Ligeti:2016npd}.

In this work, we find analytic relations between properties of tau leptons produced via $B\to D\tau\nu$ decays and corresponding observables obtained from the kinematics of visible final-state particles from the leptonic and two-body hadronic decays of the tau. This is achieved by expressing the decay rates in terms of helicity amplitudes and by analytically solving the phase-space integrals related to the kinematics of the virtual tau and final-state neutrinos. In particular, we show that the perpendicular polarization $P_\perp$~\cite{Tanaka:1994ay} and the forward-backward asymmetry $A_\tau$~\cite{Duraisamy:2013kcw} of the tau lepton can be directly extracted from an angular asymmetry of the visible decay products of the tau. This article is organized as follows. In Section~\ref{sec:production}, we introduce the tau  properties in $B\to D\tau\nu$ at the production level and discuss their kinematic features. Subsequently, in Section~\ref{sec:tautopinu}, we show how to extract the tau polarizations and forward-backward asymmetry from final-state kinematics, focusing on the full decay chains $B\to D\nu[\tau \to \{\pi\nu, \rho\nu, \ell\nu\bar{\nu}\}]$. In Section~\ref{sec:pheno}, we compare the sensitivity of the different tau decay modes to the tau properties and discuss the measurement prospects at BELLE II. We conclude in Section~\ref{sec:conclusions}.

\section{Properties of the tau lepton produced via $B\to D\tau\nu$ decays}\label{sec:production}
The differential decay rate for $B\to D^{(*)}\tau\nu$, $d\Gamma$, as well as the three polarization states of the tau lepton are defined in terms of helicity matrix elements $\mathcal{M_\pm}$ as
\begin{align}\label{eq:pol}
d\Gamma&=\frac{(2\pi)^4\,d \Phi_3}{2 m_B}\frac{|\mathcal M_+|^2+|\mathcal M_-|^2}{2}\,,&
d\Gamma dP_L&=\frac{(2\pi)^4\,d \Phi_3}{2 m_B}\left({|\mathcal M_+|^2-|\mathcal M_-|^2}\right)\,,\\\nonumber
d\Gamma dP_\perp&=\frac{(2\pi)^4\,d \Phi_3}{2 m_B}2{\rm Re}\left[\mathcal M_{+}\mathcal M_{-}^\dagger\right]\,,& d\Gamma dP_T&=\frac{(2\pi)^4\,d \Phi_3}{2 m_B}2{\rm Im}\left[\mathcal M_{+}\mathcal M_{-}^\dagger\right]\,,
\end{align}
where $d \Phi_3$ is the corresponding three-body phase space element and $dP_L$, $dP_\perp$, and $dP_T$ denote the longitudinal, perpendicular, and transversal tau polarizations, respectively. Notice that $dP_L$ and $dP_\perp$ depend on the frame in which the helicities of the tau lepton are defined. In turn, $dP_T$ points perpendicular to the $D-\tau$ decay plane and is thus invariant under boosts contained in this plane. In particular, it is invariant under boosts that connect the $B$ rest frame, the $\tau$ rest frame, and the $q$ rest frame, where $q$ is the four-momentum of the $\tau-\nu$ pair. In the absence of strong phases, a non-zero $P_T$ polarization signals violation of time-reversal symmetry.

\begin{figure}[t]
\begin{tabular}{cc}
  \includegraphics[width=8cm]{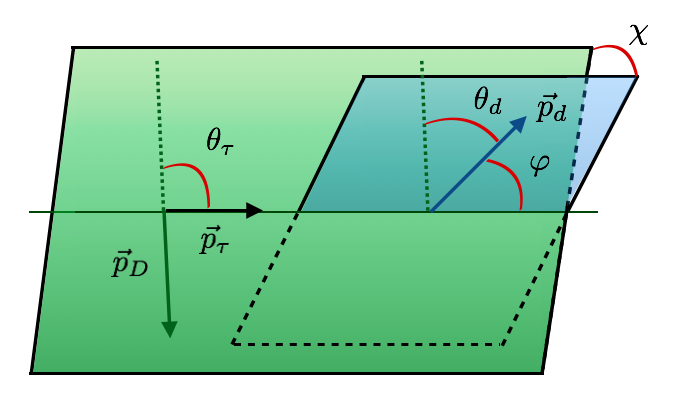}
\end{tabular}
\caption{Kinematics of the chain decay $B\to D \nu [\tau \to d \nu(\bar{\nu})]$, where $d=\{\pi,\rho,\ell\}$. The momenta of the $D$ meson and the tau lepton span the decay plane in $B\to D\tau\nu$ (in green). The momenta of the tau lepton and its visible decay product $d$ span the plane of the subsequent decay $\tau\to d\nu(\bar{\nu})$ (in blue).}
\label{fig:kinematics}
\end{figure}

In this work, we define the polarizations in the tau rest frame. The matrix element $\mathcal M_{\lambda}$ in Eq.~(\ref{eq:pol}) thus corresponds with the production of a tau lepton of helicity $\lambda=\pm1/2$ in this frame. The decay rate for a $\tau$ lepton polarized along a direction $\hat s$ is then given by~\cite{Tanaka:1994ay}
\begin{align}
d\Gamma(\hat s)=d\Gamma\big[1 + \frac{1}{2}\left(dP_L\,\hat e_\tau+dP_{\perp}\,\hat  e_\perp+dP_T\,\hat e_T \right)\cdot \hat s\big].\label{eq:Gammaspoln}
\end{align}
We choose our coordinate system $\{\hat e_\tau, \hat e_\perp,\hat e_T\}$ as
\begin{align}
\hat e_\tau&=\frac{\vec p_\tau}{|\vec p_\tau|},\qquad \hat e_T =\frac{\vec p_D\times\vec p_\tau}{|\vec p_D\times\vec p_\tau|}, \qquad \hat e_\perp =\hat e_T\times \hat e_\tau,
\label{eq:basistrans}
\end{align}
where $\vec p_{\tau}$ and $\vec p_D$ are the momenta of the $\tau$ lepton and the $D$ meson defined in the $q$ rest frame. The decay kinematics are illustrated in Fig.~\ref{fig:kinematics}.
%%%%%%%%%%%%%%%%%%%%%%%%%%%%%%%%
%%%%%%%%%%%%%%%%%%%%%%%%%%%%%%%%

\begin{figure}[t]
\centering
\includegraphics[height=2in]{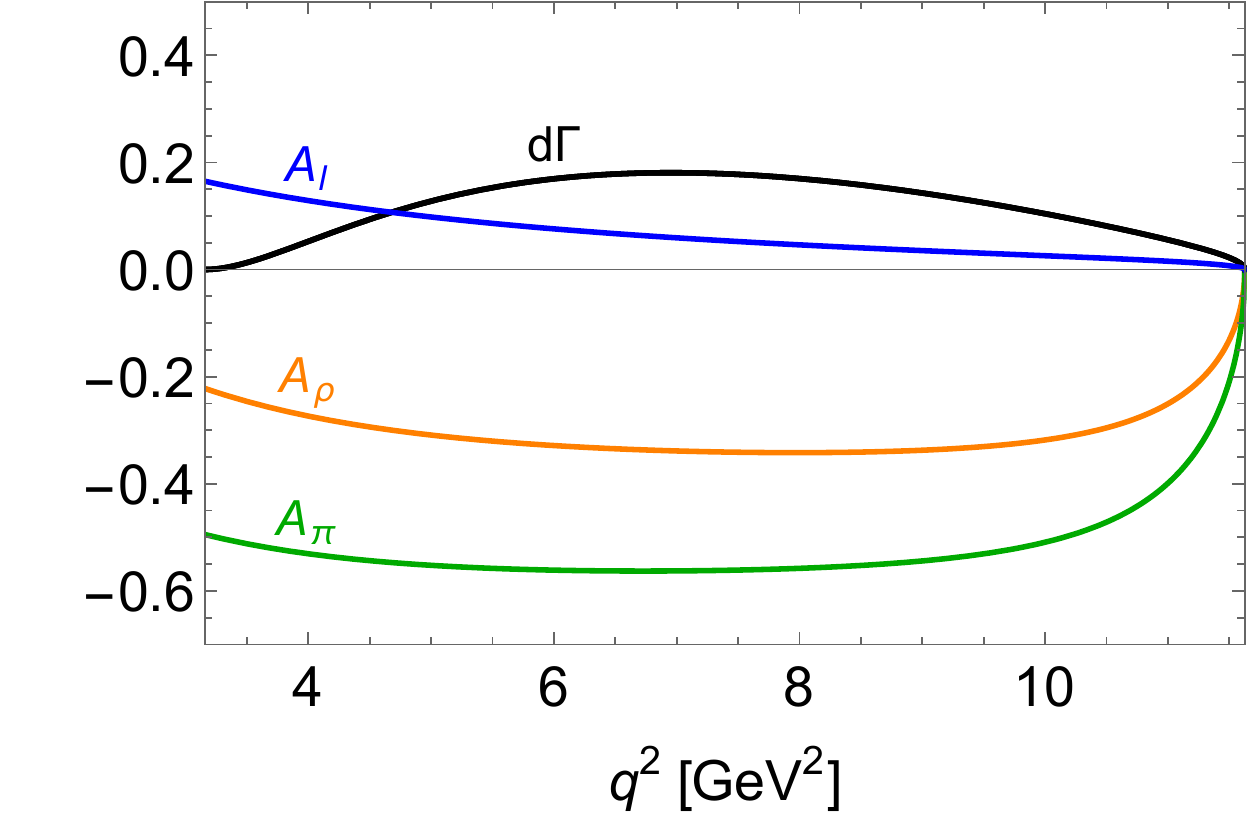} \hspace*{0.5cm} \includegraphics[height=2in]{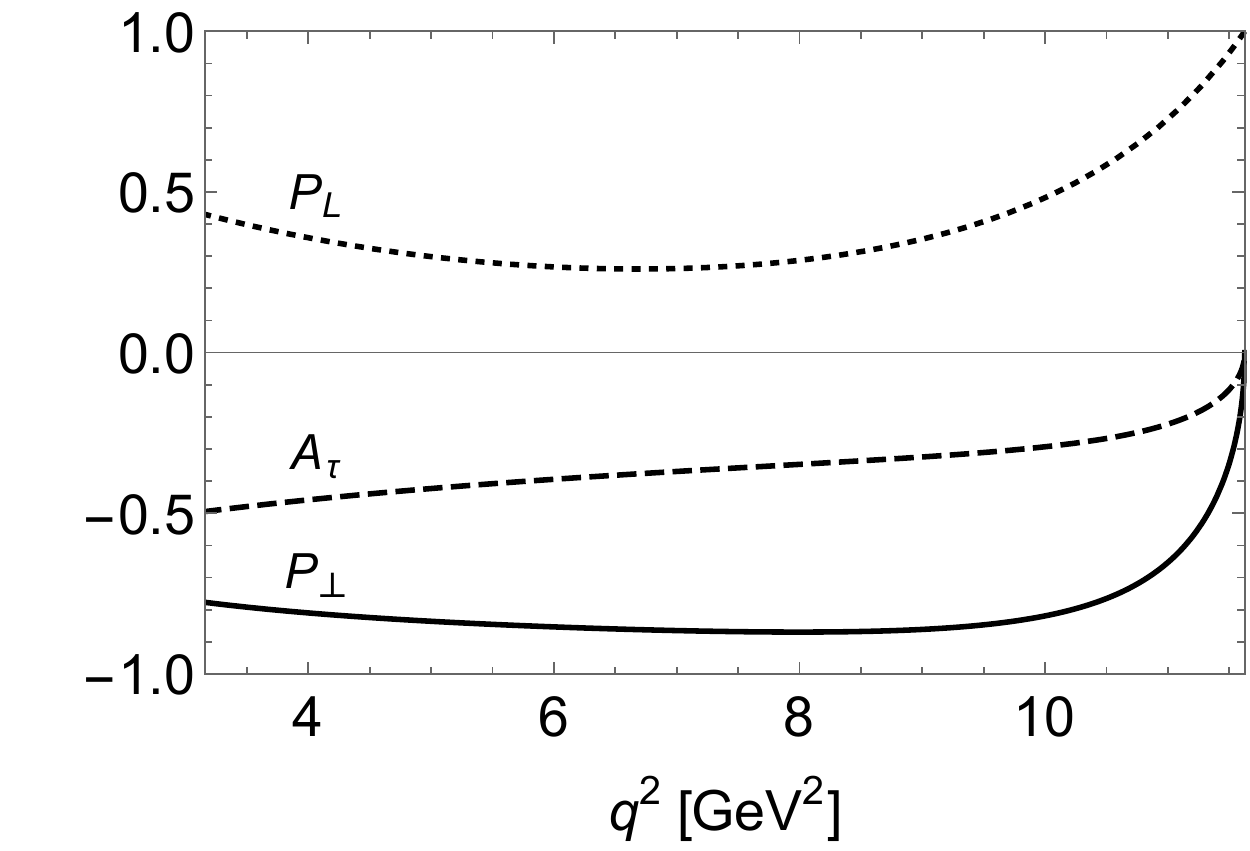}
\caption{Left: Normalized differential decay rate $(d\Gamma/dq^2)/\Gamma$ [GeV$^{-2}$] (black), and forward-backward asymmetry $A_d(q^2)$ from Eq.~(\ref{eq:dAFBpi}) for the pion (green), $\rho$ meson (orange), and charged lepton $\ell$ (blue) in $B^-\to D^0\bar \nu_\tau [\tau^-\to \pi^-\nu_\tau]$. Right: Longitudinal polarization $P_L(q^2)$ (dotted), {per\-pen\-di\-cu\-lar} polarization $P_\perp(q^2)$ (plain), and $\tau$ forward-backward asymmetry $A_{\tau}(q^2)$ (dashed) from Eq.~(\ref{eq:dAFBdq2}).}
\label{fig:pol-q2}
\end{figure}

The unpolarized differential rate depends on two kinematic variables that can be taken to be $q^2$ and the angle that the $\tau$ lepton forms with the recoil against the direction of the
$D$ in the $q$ rest frame, $\cos\theta_\tau=-\vec p_D\cdot \vec p_\tau/|\vec{p}_D||\vec{p}_\tau|$. We define the forward-backward asymmetry associated to this angle as
\begin{align}
\label{eq:dAFBdq2} \frac{d\Gamma}{dq^2}A_{\tau}(q^2) & = \int_0^1 d\cos\theta_\tau \frac{d^2\Gamma}{dq^2 d\cos\theta_\tau} - \int_{-1}^0 d\cos\theta_\tau \frac{d^2\Gamma}{dq^2 d\cos\theta_\tau}.
\end{align}
The differential rate $d\Gamma/dq^2$, normalized to the total decay rate for $B^-\to D^0 \tau^-\bar\nu_\tau$, denoted as $\Gamma$, is shown in Figure~\ref{fig:pol-q2}, left. The differential tau polarizations $P_L(q^2)$ and $P_\perp(q^2)$ are obtained by partially integrating $dP_L$ and $dP_\perp$ in Eq.~(\ref{eq:pol}) over the respective phase space. In Figure~\ref{fig:pol-q2}, right, we show $P_L(q^2)$, $P_\perp(q^2)$, and $A_\tau(q^2)$ in $B^-\to D^0\tau^-\bar{\nu}_\tau$ for the kinematic range, $m_\tau^2 \leq q^2 \leq (m_B-m_D)^2$. All three quantities are sizeable over most of the $q^2$ spectrum, which will be beneficial for a measurement. Near the endpoint $q^2=(m_B-m_D)^2$, the tau lepton recoils back-to-back against the right-handed anti-neutrino in the $B$ rest frame. Therefore, the $\tau^-$ is purely longitudinally polarized with $P_L=+1$, as can be observed in the figure. The average tau polarizations and asymmetry in the full sample of $B\to D\tau\nu$ events are given by
\begin{align}\label{eq:plav}
P_L = \frac{1}{\Gamma}\int dq^2\frac{d\Gamma}{dq^2}P_L(q^2),\quad P_\perp = \frac{1}{\Gamma}\int dq^2\frac{d\Gamma}{dq^2}P_\perp(q^2),\quad A_\tau = \frac{1}{\Gamma}\int dq^2\frac{d\Gamma}{dq^2}A_\tau(q^2).
\end{align}
Numerically, in the SM these average tau properties in $B^-\to D^0\tau^-\bar{\nu}_\tau$ amount to~\footnote{The errors quoted for these predictions are due to form factor uncertainties. Form factors have been implemented as described in Ref.~\cite{Alonso:2016gym}; $f_+(q^2)$ is obtained from fits to the measured $B\to D\ell\nu$ spectra by the Heavy Flavor Averaging Group in Ref.~\cite{Amhis:2016xyh}, whereas for the scalar form factor $f_0(q^2)$ we use the lattice QCD calculation in~\cite{Lattice:2015rga}. Our predictions confirm results found in earlier studies. In particular, we find agreement with $P_L$ in Ref.~\cite{Tanaka:2010se}, $P_\perp$ in Ref.~\cite{Ivanov:2015tru}, and $A_\tau$ in Ref.~\cite{Sakaki:2012ft}. Sign differences are due to the different choices of reference directions made in these articles.}
\begin{align}
P_L = 0.34(3),\quad P_\perp = -0.839(7),\quad A_\tau = -0.359(3).
\end{align}
It is worthwhile noting that the uncertainties for $P_\perp$ and $A_\tau$ are much smaller than for $P_L$. This is mainly due to the fact that in the SM prediction for $B\to D\tau\nu$, $P_L$ is the result of a strong cancellation between the helicity-favored ($\lambda=-1/2$) and helicity-suppressed ($\lambda= +1/2$) contributions to the rate. Only the latter depend on the ratio of form factors $(f_0(q^2)/f_+(q^2))^2$, which causes a larger overall uncertainty than in the case of $P_\perp$ and $A_\tau$.
  
By inspecting Eq.~(\ref{eq:pol}), it is apparent that the longitudinal tau polarization $P_L(q^2)$ is independent from the differential rate $d\Gamma/dq^2$, so that more information is needed to determine it unambiguously. The per\-pen\-di\-cu\-lar polarization, $P_\perp(q^2)$, which intrinsically contains information on the interference between the two tau helicity states, cannot be obtained from $d\Gamma/dq^2$. The asymmetry $A_{\tau}(q^2)$ probes the interference between the longitudinal and time-like components in the production of the $\tau-\nu$ pair and projects on tau leptons with positive helicity, $\lambda = +1/2$ (see for instance Ref.~\cite{Alonso:2016gym}). Since the angle $\theta_\tau$ cannot be reconstructed from the $\tau$ decay products, $A_{\tau}$ is not a direct observable either. In what follows, we will show that $P_L(q^2)$, $P_\perp(q^2)$, and $A_\tau(q^2)$ can be extracted with good sensitivity from kinematic distributions of the $\tau$ decay products beyond the differential decay rate $d\Gamma/dq^2$ in $B\to D\tau\nu$. 
%---------------------------------------------------------------------------
%---------------------------------------------------------------------------

\section{Observables from final-state kinematics}\label{sec:tautopinu}
In order to extract the $\tau$ polarizations directly from final-state kinematics, we need to consider the full production and decay chain of the $\tau$ lepton, i.e., $B\to D\nu[\tau \to \{\pi\nu, \rho\nu, \ell\nu\bar{\nu}\}]$. The two-body decays are particularly promising in this regard, because the meson carries more information on the $\tau$ kinematics than the lepton from the three-body decay.

The visible kinematics of the decay chain $B\to D\nu[\tau \to d\nu(\bar{\nu})]$ can be described in terms of three variables. We choose them as $q^2$, the energy of the charged particle $d$ in the $\tau$ decay, $E_{d}$, and the angle $\cos\theta_d= -\vec{p}_d\cdot \vec{p}_D/(|\vec{p}_d| |\vec{p}_D|)$, the latter two being defined in the $q$ rest frame.
The fully-differential decay rate can then be expressed as
\begin{align}\label{eq:triple-rate}
\frac{d^3\Gamma_d}{dq^2\,dE_{d}\,d\cos\theta_{d}} = \mathcal{B}_{d}\frac{\mathcal{N}}{2m_{\tau}}\big[I_0(q^2,E_{d}) + I_1(q^2,E_{d})\cos\theta_d + I_2(q^2,E_{d})\cos^2\theta_{d} \big],
\end{align}
where $\mathcal{B}_{d}$ denotes the branching ratio of the respective decay channel $\tau\to d\nu(\bar{\nu})$. Analytical formulas of the angular coefficient functions $I_{0,1,2}(q^2,E_d)$ and the normalization $\mathcal N$ can be found for $\tau\to\ell\nu\bar{\nu}$ in Ref.~\cite{Alonso:2016gym}. We have calculated the corresponding functions for $\tau\to\pi\nu$ and $\tau\to\rho\nu$ using the same methods. By integrating over $\cos\theta_{d}$, we find the double-differential rate
\begin{align}\label{eq:d2G}
\frac{d^2\Gamma_d}{dq^2\,dE_{d}} = \int_{-1}^{1} d\cos\theta_{d}\,d^3\Gamma_d = \mathcal{B}_{d}\frac{\mathcal{N}}{m_{\tau}}\big[I_0(q^2,E_{d}) + \frac{1}{3} I_2(q^2,E_{d})\big].
\end{align}
Complementary to the decay rate, we define the forward-backward asymmetry of the pion with respect to the $D$ meson,
\begin{align}\label{eq:d2AFB}
\mathcal{B}_{d}\frac{d\Gamma}{dq^2}\frac{dA_{d}}{dE_{d}} = \int_0^1 d\cos\theta_{d}\,d^3\Gamma_d - \int_{-1}^0 d\cos\theta_{d}\,d^3\Gamma_d = \mathcal{B}_{d}\frac{\mathcal{N}}{2m_{\tau}}I_1(q^2,E_{d}). 
\end{align}
Hence $d^2\Gamma_d/dq^2dE_d$ probes the angular coefficients $I_0$ and $I_2$, whereas $dA_{d}/dE_d$ is sensitive to $I_1$. The asymmetry $dA_{d}/dE_d$ is purely induced by the interference of longitudinal and time-like intermediate states of the $\tau - \nu$ pair.

Let us now focus on the dependence of these double-differential distributions. Eqs.~(\ref{eq:d2G}) and~(\ref{eq:d2AFB}), on the tau forward-backward asymmetry, $A_\tau(q^2)$, defined in Eq.~(\ref{eq:dAFBdq2}), and the differential tau polarizations $P_L(q^2)$ and $P_\perp(q^2)$. We introduce the dimensionless variables
\begin{align}
s_d = \frac{E_{d}}{\sqrt{q^2}},\quad r_d=\frac{m_d}{\sqrt{q^2}},\quad r_{\tau} = \frac{m_{\tau}}{\sqrt{q^2}}.
\end{align}
For the sake of simplicity, in what follows, we neglect the mass effects $r_\pi$ and $r_\ell$ in the decays $\tau\to\pi\nu$ and $\tau\to\ell\nu\bar{\nu}$.~\footnote{The numerical effect of this approximation on our observables is at the per mille level.} The differential decay rate can then be expressed in terms of $d\Gamma/dq^2$ and $P_L(q^2)$ as~\cite{Tanaka:2010se}
\begin{align}
\frac{d^2\Gamma_d}{dq^2\,ds_{d}} = \mathcal{B}_{d}\frac{d\Gamma}{dq^2}\Big[f_0^d + f_L^d P_L(q^2)\Big],
\end{align}
where the integrated coefficient functions satisfy
\begin{align}
\int_{s_d^{\text{min}}}^{s_d^{\text{max}}} ds_d\,f_0^d = 1\ \text{ and }\int_{s_d^{\text{min}}}^{s_d^{\text{max}}} ds_d \,f_L^d = 0,\quad \text{with } s_d^{\text{min}} = \frac{r_\tau^2}{2}\Big(1+\frac{r_d^2}{r_\tau^2}\Big),\ s_d^{\text{max}} = \frac{1}{2}\Big(1+r_d^2\Big).
\end{align}
For the pion and rho decay channels, the coefficients are given by
\begin{align}\label{eq:dG22body}
 f^\pi_0  = \frac{2}{1-r_{\tau}^2},\quad &f^\pi_L = - \frac{2(1-4s_\pi +r_{\tau}^2)}{(1-r_{\tau}^2)^2},\\
 f^\rho_0=\frac{2r_\tau^2}{\left(1-r_\tau^2\right) \left(r_\tau^2-r_\rho^2\right)},\quad&f_L^\rho=-\frac{2 r_\tau^2 \left(r_\tau^2-2 r_\rho^2\right) \left(\left(r_\tau^2+1\right) \left(r_\rho^2+r_\tau^2\right)-4 r_\tau^2 s_\rho\right)}{\left(1-r_\tau^2\right)^2 \left(r_\tau^2-r_\rho^2\right)^2 \left(2 r_\rho^2+r_\tau^2\right)}.
\end{align}
For the lepton mode, the coefficients are defined as piecewise functions, depending on the region of phase space~\cite{Alonso:2016gym},
\begin{align}\label{eq:dG2leptonic}
s_\ell   &\in\left(0,\frac{r_\tau^2}{2}\right):\ f_0^\ell =\frac{8 s_\ell^2 \left(9 r_{\tau }^4+9 r_{\tau
   }^2-8 s_\ell r_{\tau }^4-8 s_\ell r_{\tau }^2-8 s_\ell\right)}{3 r_{\tau }^6},\ f_L^\ell =\frac{8 s_\ell^2  \left(1-r_{\tau }^2\right) \left(8 s_\ell r_{\tau
   }^2+8 s_\ell-3 r_{\tau }^2\right)}{3 r_{\tau }^6},\\\nonumber
s_\ell   &\in\left(\frac{r_\tau^2}{2},\frac12\right):\ f_0^\ell =\frac{2 \left(1-2 s_\ell\right) \left(5+10 s_\ell-16 s_\ell^2\right)}{3  \left(1-r_{\tau }^2\right)},\ f_L^\ell =\frac{2 \left(1-2s_\ell\right) \left(16 s_\ell^2 r_{\tau }^2+2 s_\ell r_{\tau }^2-32 s_\ell^2+2 s_\ell+r_{\tau }^2+1\right)}{3  \left(1-r_{\tau}^2\right)^2}.
\end{align}
The energy of the visible decay particle $d$, or equivalently the variable $s_d$, is essential to extract $P_L(q^2)$ from $d^2\Gamma_d/dq^2ds_d$. Similarly, the differential forward-backward asymmetry $dA_{d}/ds_d$ can be expressed in terms of $A_{\tau}(q^2)$ and $P_{\perp}(q^2)$ as
\begin{align}\label{eq:d2AFBpi}
\frac{dA_{d}}{ds_d} & = f^d_A A_{\tau}(q^2) + f^d_{\perp}P_{\perp}(q^2),
\end{align}
where the coefficients for the pion and rho decay modes are given by
\begin{align}\label{eq:AFBfAfperp}
f^\pi_A &= -\frac{4(s_\pi - r_{\tau}^2+s_\pi r_{\tau}^2)(r_{\tau}^2-2 s_\pi)}{s_\pi(1-r_{\tau}^2)^3},\quad f^\pi_{\perp}= -\frac{8r_{\tau}(1-2s_\pi)(r_{\tau}^2-2 s_\pi)}{\pi s_\pi(1-r_{\tau}^2)^3},\nonumber\\
f^\rho_A & =-\frac{4 r_\tau^4 \left(2 r_\rho^4+r_\rho^2 (1-4 s_\rho)-r_\tau^4+2 r_\tau^2
   s_\rho\right) \left(r_\rho^2-\left(r_\tau^2+1\right) s_\rho+r_\tau^2\right)}{\left(1-r_\tau^2\right)^3 \left(r_\tau^2-r_\rho^2\right)^2 \left(2 r_\rho^2+r_\tau^2\right) \sqrt{s_\rho^2-r_\rho^2}},\nonumber\\
f^\rho_\perp & =-\frac{8 r_\tau^3 \left(r_\tau^2-2 r_\rho^2\right) \left(r_\rho^2-2 s_\rho+1\right)
   \left(r_\rho^2+r_\tau^4-2 r_\tau^2 s_\rho\right)}{\pi  \left(1-r_\tau^2\right)^3
   \left(r_\tau^2-r_\rho^2\right)^2 \left(2 r_\rho^2+r_\tau^2\right) \sqrt{s_\rho^2-r_\rho^2}},
\end{align}
and the coefficients for the lepton mode read
\begin{align}
s_\ell   &\in\left(0,\frac{r_\tau^2}{2}\right): & f_{A}^\ell&=\frac{16 s_\ell^2 \left(\left(2-4 s_\ell\right) r_{\tau }^2-1\right)}{3 r_{\tau }^4},\quad f_\perp^\ell =\frac{64 s_\ell^2 \left(r_{\tau }^2-2 s_\ell \left(r_{\tau }^2+1\right)\right)}{3 \pi 
   r_{\tau }^5},\\\nonumber
s_\ell   &\in\left(\frac{r_\tau^2}{2},\frac12\right): & f_{A}^\ell&=\frac{4 \left(1-2 s_\ell\right)^2 \left(4 s_\ell^2 \left(r_{\tau }^4-3 r_{\tau
   }^2+3\right)+s_\ell \left(2 r_{\tau }^4-5 r_{\tau }^2+3\right)+r_{\tau }^2 \left(r_{\tau}^2-2\right)\right)}{3 s_\ell \left(1-r_{\tau }^2\right)^3},\\\nonumber
&  & f_\perp^\ell&=\frac{8 \left(1-2 s_\ell\right)^2 r_{\tau } \left(4 s_\ell^2 \left(r_{\tau }^2-2\right)+2
   s_\ell \left(r_{\tau }^2-1\right)+r_{\tau }^2\right)}{3 \pi  s_\ell \left(1-r_{\tau}^2\right)^3}.
\end{align}
Since $A_\tau$ probes purely longitudinally polarized tau leptons (see Section~\ref{sec:production}), $A_\tau(q^2)$ and $P_\perp(q^2)$ are clearly independent quantities. They can be extracted from a two-dimensional fit to the energy distribution of the forward-backward asymmetry, $dA_{d}/ds_d$. For later convenience, we also define the $s_d$-integrated  asymmetries,
\begin{align}\label{eq:dAFBpi}
A_{d}(q^2) & = F^d_A A_{\tau}(q^2) + F^d_\perp P_\perp(q^2),
\end{align}
with the integrated coefficient functions
\begin{align}
F^d_A & = \int_{s_d^{\text{min}}}^{s_d^{\text{max}}} ds_d\,f^d_A,\quad F^d_\perp= \int_{s_d^{\text{min}}}^{s_d^{\text{max}}} ds_d\,f^d_\perp.
\end{align}

The distributions $A_d(q^2)$ are shown in Figure~\ref{fig:pol-q2}, left. At the endpoint, where the $D$ meson is produced at rest, $A_d(q^2)$ tends to zero, since both $A_{\tau}(q^2)$ and $P_\perp(q^2)$ vanish (see Figure~\ref{fig:pol-q2}, right). Otherwise, the asymmetries for the pion and rho modes are sizeable over the remaining $q^2$ range. The total asymmetries are given by~\footnote{The inclusive pion and rho asymmetries, $A_\pi$ and $A_\rho$, have previously been suggested as discriminators between various contributions of new physics to $B\to D^{(\ast)}\tau\nu$ decays~\cite{Sakaki:2012ft}. Our results agree with those from Ref.~\cite{Sakaki:2012ft}, the sign difference being due to different definitions of the angle $\theta_d$.}
\begin{align}
A_\pi = -0.54,\quad A_\rho = -0.32,\quad A_\ell = +0.06.
\end{align}
The magnitude of $A_\pi$ and $A_\rho$ suggests that they will be sensitive observables of $P_\perp$ and $A_\tau$. We will quantify this fact in what follows.

\section{Phenomenology and observation prospects at BELLE II}\label{sec:pheno}
After having determined the analytic relations between visible final-state kinematics and the $\tau$ properties in $B\to D \tau\nu$ decays, we now seek to quantify the sensitivity of the differential rate, $d^2\Gamma_d/dq^2ds_d$, to $P_L(q^2)$, and of the  angular asymmetry, $dA_d/ds_d$, to $P_\perp(q^2)$ and $A_{\tau}(q^2)$. The energy
 of the visible decay particle $d$, $s_d$, will serve as a $\tau$ polarimeter for the observables. For the longitudinal polarization, a similar analysis has previously been performed in Ref.~\cite{Tanaka:2010se} for the decays $\tau\to\pi\nu$ and $\tau\to\ell\nu\bar{\nu}$. Here we extend the analysis of $d^2\Gamma_d/dq^2ds_d$ by the decay $\tau\to\rho\nu$ and present new results for the asymmetry $dA_d/ds_d$ in all three decay modes $\tau\to\pi\nu$, $\tau\to\rho\nu$, and $\tau\to\ell\nu\bar{\nu}$.
%---------------------------------------------------------------------------
\begin{figure}[t]
\centering
\includegraphics[height=2.21in]{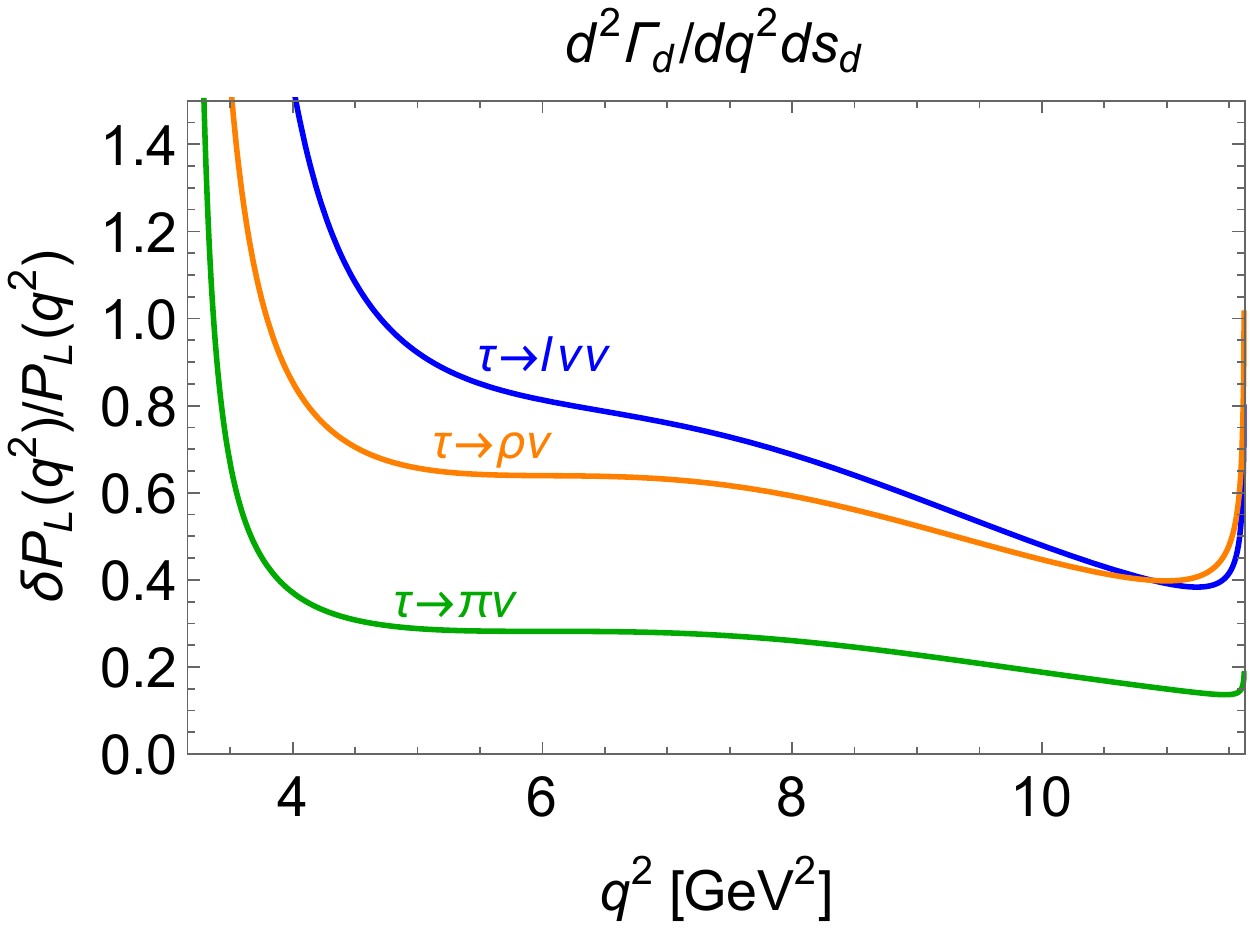} \hspace*{0.5cm} \includegraphics[height=2.17in]{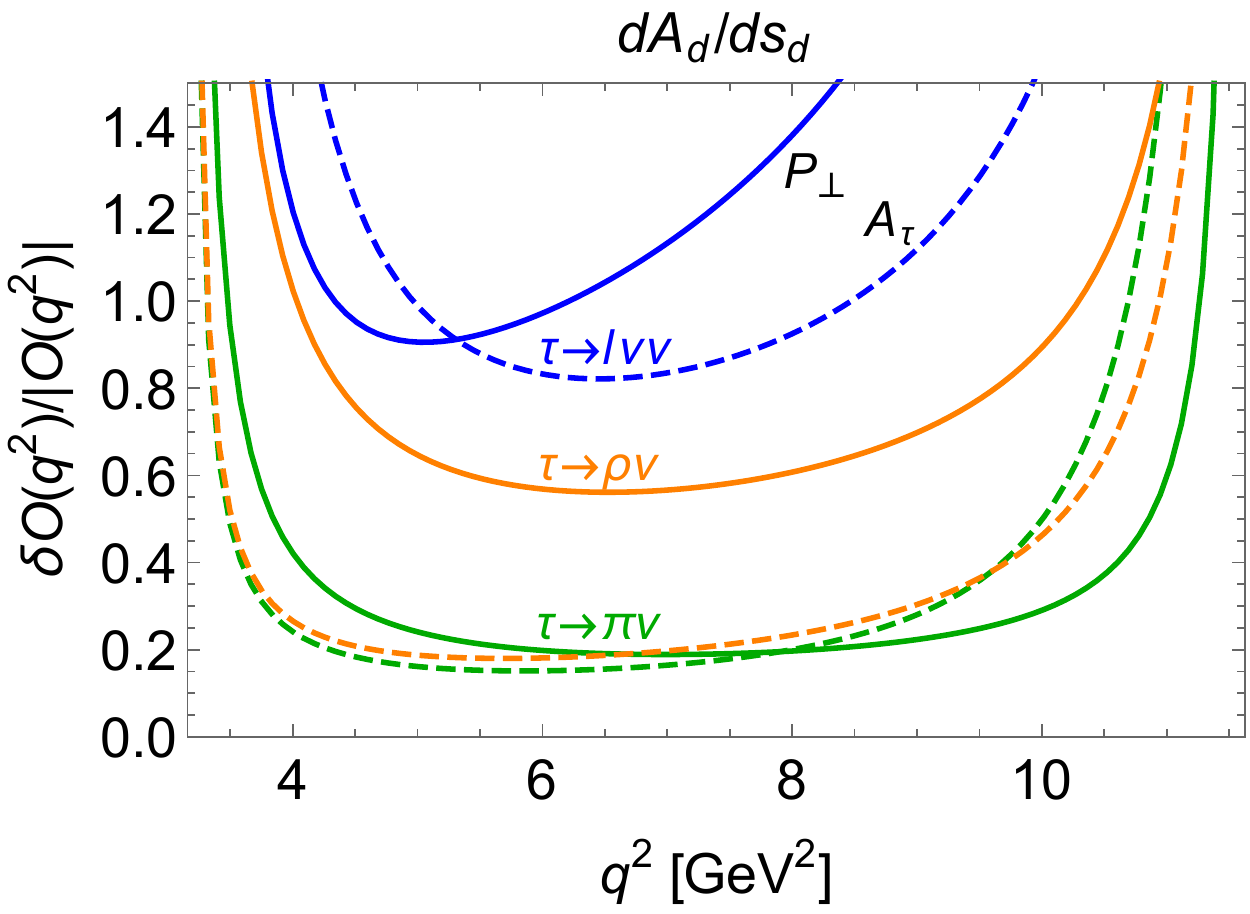}
\caption{Relative statistical uncertainties on the longitudinal polarization, $P_L(q^2)$, measured in the differential rate $d^2\Gamma_d/dq^2ds_d$ (left), and on the perpendicular polarization, $P_\perp(q^2)$ (plain), and angular asymmetry, $A_\tau(q^2)$ (dashed), measured in the asymmetry $dA_d/ds_d$ (right). Shown are the three decay channels $\tau\to\pi\nu$ (green), $\tau\to\rho\nu$ (orange), and $\tau\to\ell\nu\bar{\nu}$ (blue). Projections are for BELLE II with a total of $N=3000$ events per channel $(50\,\text{ab}^{-1})$ of data.}
\label{fig:dPLrel}
\end{figure}
%---------------------------------------------------------------------------

Let us first consider the longitudinal polarization of the tau, $P_L$. Assuming an ideal experiment with unlimited resolution in $q^2$ and $E_d$, we define the statistical uncertainty $\delta P_L(q^2)$ of measuring the longitudinal polarization in the differential rate $d^2\Gamma_d/dq^2ds_d$ as
\begin{align}
\delta P_L(q^2) = \frac{1}{\sqrt{N(q^2)}S_L(q^2)}.
\end{align}
Here $N(q^2)$ is the number of events $i=1,\dots N(q^2)$ with energy $s_d^i$ for a fixed momentum $q^2$. For a large data sample $N(q^2)$, the sensitivity $S_L(q^2)$ is given by (cf. Refs.~\cite{Davier:1992nw,Tanaka:2010se})
\begin{align}
S_L^2(q^2) & = \int ds_d \frac{f_L(r_\tau,s_d)^2}{f_0(r_\tau) + f_L(r_\tau,s_d)P_L(q^2)}.
\end{align}
In Figure~\ref{fig:dPLrel}, left, we show the relative statistical uncertainty, $\delta P_L(q^2)/P_L(q^2)$, in $d^2\Gamma_d/dq^2ds_d$ from $B^-\to D^0 \tau^-\bar\nu_\tau$ for the three decay modes $\{\tau^-\to \pi^-\nu_\tau,\tau^-\to \rho^-\nu_\tau,\tau^-\to\ell^-\bar{\nu}_\ell\nu_\tau\}$ as expected at BELLE II. Assuming the same detector performance as for BELLE I, the expected total number of events is roughly the same in each decay channel. For a luminosity of $50\,(5)\,\text{ab}^{-1}$, we expect $N\approx 3000\,(300)$ events per decay mode~\cite{Aushev:2010bq}. In all three channels, the statistical sensitivity reaches its maximum near the kinematic endpoint at large $q^2$. A precise measurement of the energy of the visible decay product in this region will thus facilitate the
 extraction of the longitudinal tau polarization to a good accuracy. By comparing the different tau decay modes, it is apparent that the pion in $\tau\to\pi\nu$ (green) has the best analyzing power, since the pion kinematics translate directly into the polarization of the tau lepton. In $\tau\to\rho\nu$ (orange), the sensitivity is reduced due to the additional decay into transversely polarized rho mesons. In $\tau\to\ell\nu\bar{\nu}$ (blue), the relation between the final-state lepton and the tau polarization is washed out by the second invisible neutrino. From the theory point of view, the decay $\tau\to\pi\nu$ is therefore the preferred channel to observe the longitudinal tau polarization.

In the full sample of $N$ events for $B^-\to D^0\tau^-\bar{\nu}_\tau$, the statistical uncertainty on the average longitudinal polarization $P_L$ from Eq.~(\ref{eq:plav}) is given by
\begin{align}\label{eq:plerror}
\delta P_L = \frac{1}{\sqrt{N}S_L},\quad\text{with } S_L^{-2} = \frac{1}{\Gamma}\int dq^2 \frac{d\Gamma}{dq^2}S_L^{-2}(q^2).
\end{align}
In Table~\ref{tab:pl}, we compare the relative statistical uncertainty, $\delta P_L/P_L$, in $B^-\to D^0 \tau^-\bar\nu_\tau$ for the decays $\tau\to \pi\nu$, $\tau\to\rho\nu$, and $\tau\to\ell\nu\bar{\nu}$. As expected from Fig.~\ref{fig:dPLrel}, left, the decay mode $\tau\to\pi\nu$ has the best overall sensitivity to the longitudinal tau polarization. Already with the complete data set collected at BELLE I, $P_L$ can be measured up to a statistical uncertainty of $\delta P_L/P_L=21\%$, which will be reduced to the $3\%$ level by the end of BELLE II. In the long term, also $\tau\to\rho\nu$ and $\tau\to\ell\nu\bar{\nu}$ will be promising decay modes to observe $P_L$ with less than $10\%$ statistical uncertainty.
%---------------------------------------------------------------------------
\begin{table}[!tb]
\centering
\renewcommand{\arraystretch}{1.5}
\begin{tabular}{|c|c|c|c|}
\hline
 & BELLE I [total] & BELLE II [1 year] & BELLE II [total] \\
\hline
$\mathcal{L}\ [\text{ab}^{-1}]/N\,\text{[events]}$ & $1/60$ & $5/300$ & $50/3000$ \\
\hline
$\delta P_L/P_L$ & $\{0.21,0.49,0.62\}$ & $\{0.10,0.22,0.28\}$ & $\{0.03,0.07,0.09\}$ \\
$\delta P_\perp/|P_\perp|$ & $\{0.62,1.8,4.0\}$ & $\{0.28,0.81,1.8\}$ & $\{0.09,0.25,0.57\}$ \\
$\delta A_\tau/|A_\tau|$ & $\{0.74,0.69,2.8\}$ & $\{0.33,0.31,1.3\}$ & $\{0.11,0.10,0.40\}$ \\
\hline
\end{tabular}
\caption{Relative statistical uncertainties on the $\tau$ polarizations, $P_L$ and $P_\perp$, and angular asymmetry, $A_\tau$, in $B^-\to D^0 \tau^-\bar\nu_\tau$ for different $\tau$ decays $\{\tau\to\pi\nu,\tau\to\rho\nu,\tau\to\ell\nu\bar{\nu}\}$. Predictions are given for the full data set from BELLE I and projections for BELLE II.}
\label{tab:pl}
\end{table}
%---------------------------------------------------------------------------

A first measurement of the longitudinal $\tau$ polarization in $B\to D^\ast \tau \nu$, with hadronic decays $\tau\to\pi\nu$ and $\tau\to\rho\nu$, has recently been performed by the BELLE collaboration~\cite{Hirose:2016wfn}. As in our approach, BELLE measures the quantities $q^2$ and $E_d$, which determine the pion or rho scattering angle against the $\tau$ direction in the $q^2$ frame, $\cos\theta_{\tau d}$. The helicity angle, $\cos\theta_{\rm hel}$, which is sensitive to the polarization in the $\tau$ rest frame, $P_L$, is then obtained by boosting the event to a pseudo $\tau$ rest frame on a cone around the $d$ direction. The sensitivity to $P_L$ obtained through this procedure is the same as in our distribution $d^2\Gamma_d/dq^2 dE_d$ from Eq.~(\ref{eq:triple-rate}). We therefore suggest to directly extract $P_L$ from the energy distribution of the visible decay particle $d$, as has been pointed out earlier in Ref.~\cite{Tanaka:2010se}.\footnote{The approach taken in Ref.~\cite{Ivanov:2017mrj} is the same as in the BELLE measurement; the angle $\theta_d$ in the former corresponds to the angle $\theta_{\rm hel}$ in the latter.}

\begin{figure}[t]
\centering
\includegraphics[height=1.78in]{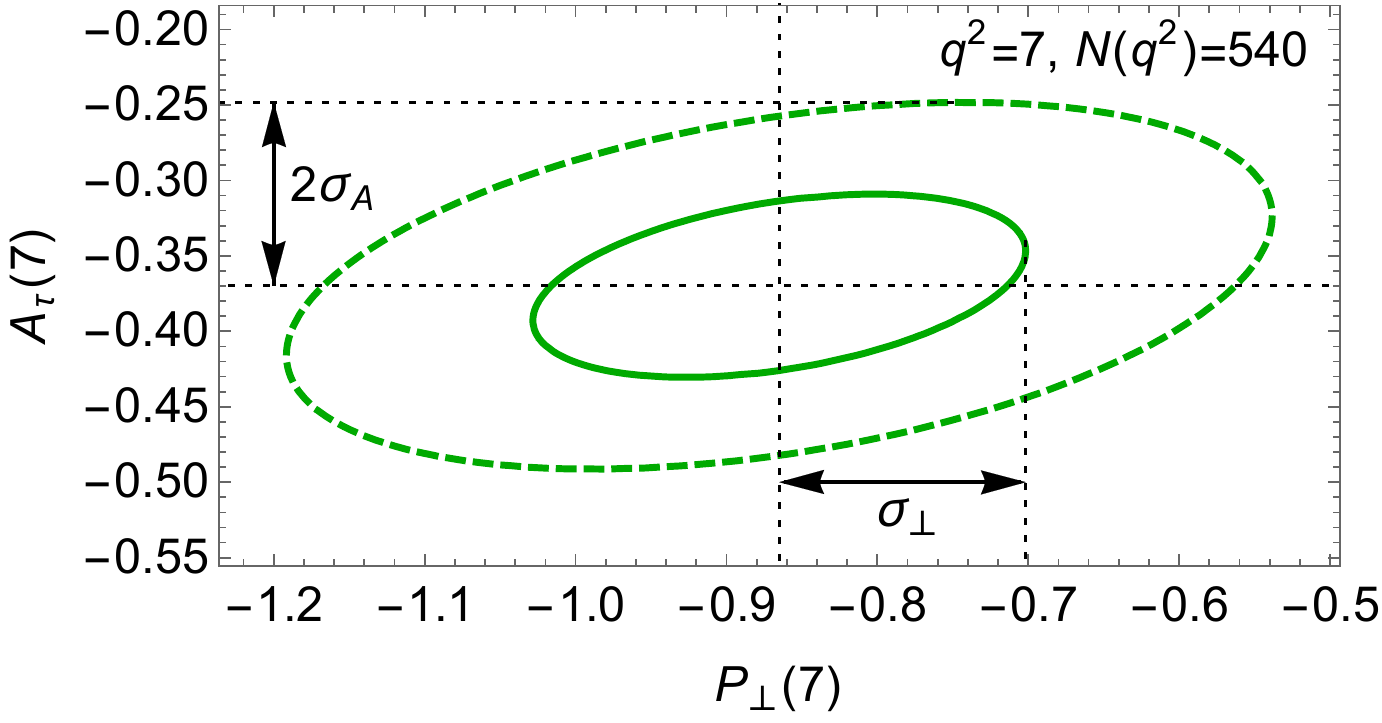}\hspace*{0.5cm}\includegraphics[height=1.8in]{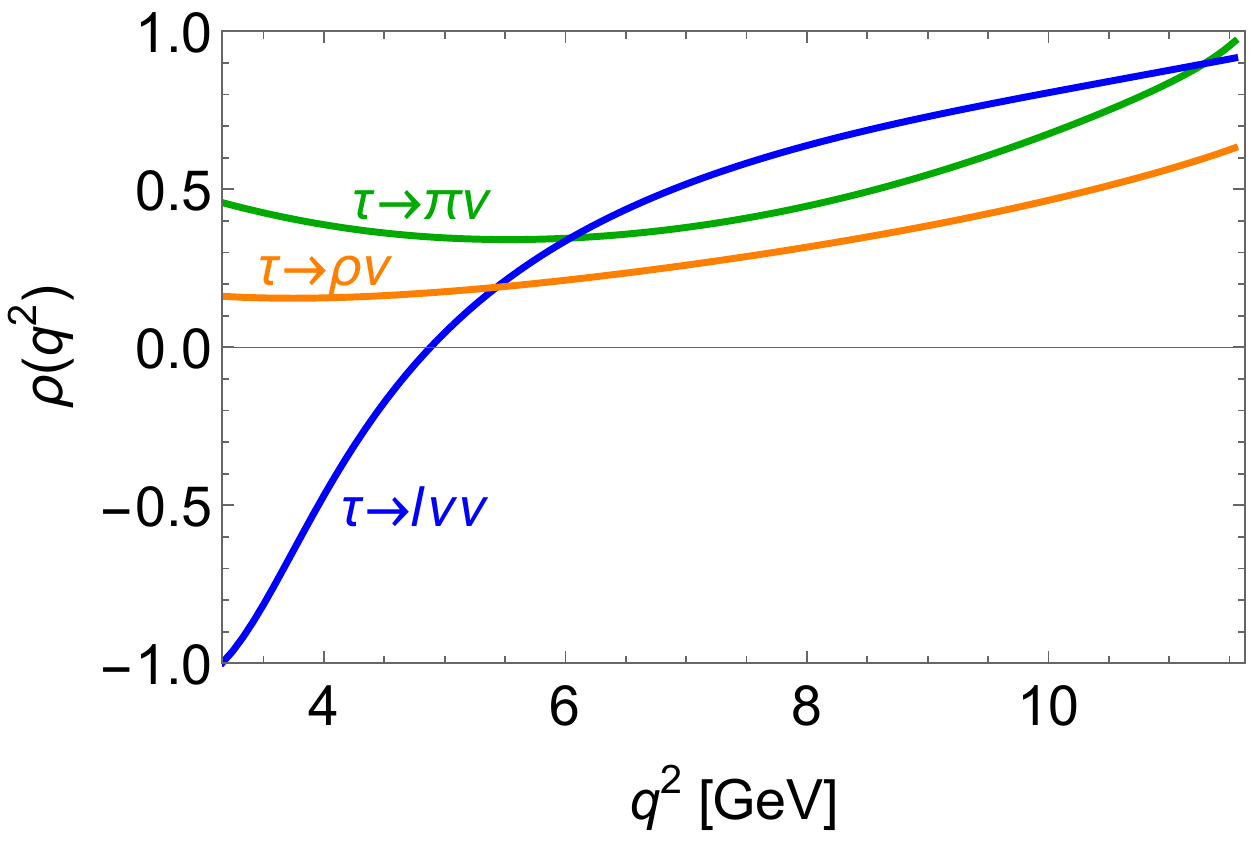}
\caption{Left: Statistical uncertainty ellipse expected from an unbinned maximum likelihood fit of $P_\perp(q^2)$ and $A_{\tau}(q^2)$ in the asymmetry $dA_\pi/ds_\pi$ of $B^-\to D^0\bar\nu_\tau[\tau^-\to\pi^-\nu_\tau]$ to $N(q^2) = 540$ events at $q^2=7$. The long dotted vertical and horizontal lines indicate the values of the estimators $\widehat{P}_\perp(q^2)$ and $\widehat{A}_{\tau}(q^2)$, equal to their standard-model predictions. The contours of the plain and dashed green ellipses lie $1$ and $2$ standard deviations away from the estimators, respectively. Right: Correlation $\rho(q^2)$ between $\widehat{P}_\perp(q^2)$ and $\widehat{A}_{\tau}(q^2)$ in $dA_d/ds_d$ for the three decay modes $\tau\to\pi\nu$ (green), $\tau\to\rho\nu$ (orange) and $\tau\to\ell\nu\bar{\nu}$ (blue).}
\label{fig:ellipse}
\end{figure}

To extract $P_\perp(q^2)$ and $A_{\tau}(q^2)$ from the forward-backward asymmetry $A_d$, we propose an unbinned maxi\-mum likelihood fit to the energy distribution $dA_d/ds_d$ from Eq.~(\ref{eq:d2AFBpi}). We define the probabilities $\mathcal{P}_+(q^2,s_d)$ and $\mathcal{P}_-(q^2,s_d)$ to find an event with decay particle energy $s_d$ and $\cos\theta_d > 0$ or $\cos\theta_d < 0$, respectively, for a given $q^2$ (bin) as
\begin{align}
\mathcal{P}_+(q^2,s_d) & = \frac{f_0(r_\tau) + f_L(r_\tau,s_d)P_L(q^2) + f_A(r_\tau,s_d)A_{\tau}(q^2) + f_\perp(r_\tau,s_d)P_\perp(q^2)}{1+F_A(r_\tau)A_{\tau}(q^2) + F_\perp(r_\tau) P_\perp(q^2)},\\
\mathcal{P}_-(q^2,s_d) & = \frac{f_0(r_\tau) + f_L(r_\tau,s_d)P_L(q^2) - f_A(r_\tau,s_d)A_{\tau}(q^2) - f_\perp(r_\tau,s_d)P_\perp(q^2)}{1 - F_A(r_\tau)A_{\tau}(q^2) - F_\perp(r_\tau) P_\perp(q^2)}.
\end{align}
For a data set of $N(q^2)$ events from the decay $B\to D\nu[\tau\to d\nu(\bar{\nu})]$, the log likelihood function of the variable $s_d$ with the parameters $P_\perp(q^2)$ and $A_{\tau}(q^2)$ is given by
\begin{align}
\ln\mathcal{L}(s_d\vert\{P_\perp(q^2),\,A_{\tau}(q^2)\}) & = \sum_{i=1}^{N_+(q^2)} \ln(\mathcal{P}_+(q^2,s_d^i)) + \sum_{j=1}^{N_-(q^2)}\ln(\mathcal{P}_-(q^2,s_d^j)),\\\nonumber
\text{where } N_{\pm}(q^2) & = N(q^2)\frac{1\pm A_d(q^2)}{2}.
\end{align}
For large $N(q^2)$, the likelihood function is Gaussian distributed around the estimators $\widehat{P}_\perp(q^2)$ and $\widehat{A}_{\tau}(q^2)$ which maximize $\ln\mathcal{L}$. The covariance matrix is then obtained from the second derivatives of the log likelihood,
\begin{align}
\text{cov}(\widehat{P}_\perp(q^2),\,\widehat{A}_{\tau}(q^2)) = 
\begin{pmatrix}
\sigma_\perp^2 & \rho\,\sigma_\perp\sigma_A\\
\rho\,\sigma_\perp\sigma_A & \sigma_A^2
\end{pmatrix},
\quad \text{with } \text{cov}^{-1}(\widehat{V}_1,\widehat{V}_2)_{ij} = -\frac{\partial^2\ln\mathcal{L}}{\partial V_i\,\partial V_j}\vert_{V_1=\widehat{V}_1,V_2=\widehat{V}_2}.
\end{align}
Here $\sigma_\perp$ and $\sigma_A$ are the standard deviations of the maximum likelihood estimators in a data set of $N(q^2)$ events. The correlation coefficient for $P_\perp(q^2)$ and $A_{\tau}(q^2)$ is denoted as $\rho$. The parameter pairs $\{P_\perp(q^2),A_{\tau}(q^2)\}$, which are $n$ standard deviations away from the estimators, lie on an ellipse defined by
\begin{align}
\ln\mathcal{L}(s_d\vert \{\widehat{P}_\perp\pm n\sigma_\perp,\,\widehat{A}_{\tau}\pm n\sigma_A\}) - \ln\mathcal{L}(s_d\vert \{\widehat{P}_\perp,\,\widehat{A}_{\tau}\}) = -\frac{n^2}{2}.
\end{align}
In Figure~\ref{fig:ellipse}, left, we display this ellipse for the decay mode $\tau\to\pi\nu$. We choose $q^2=7$, where $d\Gamma/dq^2$ and $A_\pi(q^2)$ reach their maximum (see Figure~\ref{fig:pol-q2}, left). The data sample in our example comprises $N(q^2=7)=540$ events, corresponding to a total data set of $N=3000$ events expected at BELLE II with $50\,\text{ab}^{-1}$ of data luminosity. We have assumed that the estimators are equal to their standard-model expectations, $\widehat{P}_{\perp}(7)=-0.86$ and $\widehat{A}_{\tau}(7)=-0.37$. Shown are the contours $n=1$ (plain) and $n=2$ (dashed), corresponding to one and two standard deviations from the estimators.  Since the standard deviations for $A_{\tau}(7)$ and $P_\perp(7)$ are comparable in magnitude, we expect that $A_{\tau}$ and $P_L$ can be extracted with similar precision from a given data set. The tilt of the ellipse indicates the correlation $\rho(7)=0.38$ between $P_\perp(7)$ and $A_{\tau}(7)$. In Figure~\ref{fig:ellipse}, right, the correlation coefficient $\rho(q^2)$ between $\widehat{P}_\perp(q^2)$ and $\widehat{A}_{\tau}(q^2)$ is shown for all three tau decay modes. In the hadronic decays, the correlation is moderate at low to intermediate $q^2$ and in particular around $q^2=7$, where most of the events are expected. In this region, $E_d$ is thus a good discriminator between $P_\perp$ and $A_\tau$. In the leptonic modes, the correlation between $P_\perp$ and $A_\tau$ is different, which could, in principle, allow for an independent extraction of these quantities by combining results from different tau decay modes.
%---------------------------------------------------------------------------
%---------------------------------------------------------------------------

To estimate the sensitivity of the angular asymmetry to $P_\perp(q^2)$ and $A_\tau(q^2)$ individually, we proceed as for $P_L(q^2)$. The statistical uncertainties are given by the standard deviations as
\begin{align}
\delta P_\perp (q^2) = \frac{1}{\sqrt{N(q^2)}S_\perp(q^2)} = \sigma_\perp(q^2),\qquad \delta A_\tau (q^2) = \frac{1}{\sqrt{N(q^2)}S_A(q^2)} = \sigma_A(q^2).
\end{align}
In Figure~\ref{fig:dPLrel}, right, we show the relative statistical uncertainties for $P_\perp(q^2)$ (plain) and $A_\tau(q^2)$ (dashed) as expected from a maximum likelihood fit of $dA_d/ds_d$ to BELLE-II data corresponding to $50\,\text{ab}^{-1}$ luminosity. In each of the three tau decay modes, the uncertainties are smallest in the region of intermediate $q^2$, where the decay rate is high. As in the case of the longitudinal polarization, the pion from the decay $\tau\to\pi\nu$ has the highest analyzing power. In this decay mode, $P_\perp(q^2)$ and $A_\tau(q^2)$ can be extracted from $dA_\pi/ds_\pi$ with a precision that is similar to $P_L(q^2)$ extracted from $d^2\Gamma_\pi/dq^2ds_\pi$ over the range of intermediate $q^2$ (cf. Figure~\ref{fig:dPLrel}, left). Interestingly, the decay mode $\tau\to\rho\nu$ can compete with $\tau\to\pi\nu$ in its sensitivity to $A_\tau$ (green and orange dashed lines in Figure~\ref{fig:dPLrel}, right). Since $A_\tau$ probes the tau polarization along the tau momentum, only the longitudinal component of the rho meson contributes. This component has the same analyzing power as the pion (see also Ref.~\cite{Ivanov:2017mrj}). The small difference in sensitivity between $\tau\to\rho\nu$ and $\tau\to\pi\nu$ is due to the meson mass effects. The decay $\tau\to\ell\nu\bar{\nu}$ is much less sensitive to $P_\perp$ and $A_\tau$ and a significant increase in statistics, such as the one that could be provided by the LHCb, would be necessary to make this mode competitive with the hadronic ones.

The statistical uncertainties on the average perpendicular polarization, $\delta P_\perp$, and tau asymmetry, $\delta A_\tau$, are defined as for the longitudinal polarization in Eq.~(\ref{eq:plerror}), with $P_L$ replaced by $P_\perp$ or $A_\tau$, respectively. The expected accuracy for a measurement of $P_\perp$ and $A_\tau$ at BELLE is shown in Table~\ref{tab:pl}. For $\tau\to\pi\nu$, the sensitivities to $P_\perp$ and $A_\tau$ are comparable, while for $\tau\to\rho\nu$ and $\tau\to\ell\nu\bar{\nu}$ the sensitivity to $A_\tau$ is higher than for $P_\perp$. While the current statistical sensitivity of BELLE I is limited to $\delta P_\perp/P_\perp\approx 60\%$ and $\delta A_{\tau}/A_{\tau}\approx 70\%$, it will improve significantly with the larger data set expected at BELLE II. In the preferred decay mode $\tau\to\pi\nu$, $P_\perp$ and $A_\tau$ are expected to be accessible with a precision of $9\%$ and $11\%$, respectively. Remarkably, the decay $\tau\to\rho\nu$ serves as an alternative channel to observe $A_\tau$ at the $10\%$ level. With hadronic tau decays, the statistical sensitivity of the asymmetry $A_d$ to $P_\perp$ and $A_\tau$ is thus not much lower than for $P_L$. The differential decay rate and the hadron asymmetry from hadronic tau decays are thus complementary observables of the tau properties in $B\to D\tau\nu$.

\section{Conclusions}\label{sec:conclusions}
In this work, we have established explicit analytical relations between the tau properties in $B\to D\tau\nu$ decays and the kinematics of visible final-state particles. For the three dominant decay modes $\tau\to\pi\nu$, $\tau\to\rho\nu$, and $\tau\to\ell\nu\bar{\nu}$, it was shown how the longitudinal tau polarization, $P_L$, can be obtained from the energy distribution of the charged decay particle $d$ in the full decay rate. Complementary to the decay rate, the angular asymmetry of $d$ against the $D$ meson direction allows us to extract the perpendicular polarization, $P_\perp$, and forward-backward asymmetry, $A_\tau$, of the tau lepton. These results provide a sound framework to gain the maximal available information on the production mechanism of the tau lepton directly from its visible decay products. The benefit of this approach is that the partial reconstruction of the tau rest frame can be avoided, so that the interpretation of the final state in terms of tau properties is immediate and transparent.

To quantify our results, we have performed a numerical statistics analysis for the BELLE and BELLE II experiments. Among the three considered decay modes, $\tau\to\pi\nu$ is shown by our analysis to be the most sensitive channel to all three tau properties $P_L$, $P_\perp$, and $A_\tau$, because the kinematics of the scalar pion directly reflect the tau helicity state. With the full data set obtained at BELLE, we expect a relative statistical precision of $\delta P_L/P_L=21\%$, $\delta P_\perp/|P_\perp| = 62\%$, and $\delta A_\tau/|A_\tau| = 74\%$. At BELLE II, with $50$~ab$^{-1}$ of data, the sensitivity is significantly improved, yielding an ultimate statistical precision of $\delta P_L/P_L=3\%$, $\delta P_\perp/|P_\perp| = 9\%$, and $\delta A_\tau/|A_\tau| = 11\%$. The decay $\tau\to\rho\nu$ has a comparable sensitivity to the asymmetry $A_\tau$, which imprints itself only on the longitudinal component of the vector meson rho.We therefore strongly encourage experimentalists to continue and intensify the investigation of two-body hadronic tau decays. In the leptonic decay $\tau\to\ell\nu\bar{\nu}$, the access to the tau properties is washed out by the presence of the second neutrino, making them much less sensitive. Nonetheless, this could be compensated by sheer statistics if the kinematic distributions were measured at LHCb. Alternative decay modes, such as the three-prong tau decay into three pions, might be similarly sensitive to tau properties.

The goal of our paper was to outline the strategy to observe tau properties $P_L$, $P_\perp$, and $A_\tau$ from final-state kinematics with the application to $B\to D\tau\nu$ decays. A similar analysis for $B\to D^\ast\tau\nu$ is an interesting extension of this framework, which we leave for future work. Beyond the standard model, the investigation of the tau properties along the same lines will provide us with valuable information on a possible modification of the production process. The results can shed light on the apparent discrepancy between the SM predictions and measurements of semi-leptonic $B$ decays with taus.

\section{Acknowledgments}
We thank Karol Adamczyk and Maria Rozanska for discussions of experimental aspects. JMC wants to acknowledge Victoria for her unconditional support during the last stages of this work. SW acknowledges funding by the Carl Zeiss Foundation through a \emph{Junior-Stiftungsprofessur}.

\bibliography{BtoDtaunu.bib}

\end{document}